\documentclass[sort&compress,10pt,twocolumn,final,3p]{elsarticle}

\usepackage[utf8]{inputenc}
\usepackage{hyperref}
\usepackage[]{graphicx}
\usepackage{subfigure}
\usepackage{array}
\usepackage{multirow}
\usepackage{amsmath,amssymb}
\usepackage{bm}	
\usepackage{xspace} 

\usepackage{color}
\usepackage{miller}

\newcommand{\etal}{\textit{et al}.\@\xspace}
\newcommand{\ie}{\textit{i.e.}\@\xspace}

\newcommand{\cf}{\textit{cf}.\@\xspace}
\newcommand{\pwscf}{\textsc{Pwscf}\@\xspace}

\newcommand{\abinitio}{\textit{ab initio}\@\xspace}
\newcommand{\Abinitio}{\textit{Ab initio}\@\xspace}

\journal{Acta Materialia}

\begin{document}

\begin{frontmatter}

\title{Oxygen - dislocation interaction in zirconium from first principles}

\author[SRMP]{Nermine Chaari\fnref{AREVA}}
\author[ILM]{David Rodney}
\author[SRMP]{Emmanuel Clouet\corref{CA}}
\cortext[CA]{Corresponding author}
\ead{emmanuel.clouet@cea.fr}
\address[SRMP]{DEN-Service de Recherches de Métallurgie Physique, CEA, Université Paris-Saclay, F-91191 Gif-sur-Yvette, France}
\address[ILM]{Institut Lumière Matière, CNRS-Université Claude Bernard Lyon 1, F-69622 Villeurbanne, France}
\fntext[AREVA]{Present address: AREVA NP, 10 rue Juliette Récamier, F-69006 Lyon, France}

\begin{abstract}
Plasticity in zirconium alloys is mainly controlled by the interaction of $1/3\,\hkl<1-210>$ screw dislocations
with oxygen atoms in interstitial octahedral sites of the hexagonal close-packed lattice. This process is studied here using \abinitio calculations based on the density functional theory.  
The atomic simulations show that a strong repulsion exists only when the O atoms lie in the dislocation core
and belong to the prismatic dislocation habit plane.  This is a consequence of the destruction of the octahedral sites 
by the stacking fault arising from the dislocation dissociation.
Because of the repulsion, the dislocation partially cross-slips to an adjacent prismatic plane, 
in agreement with experiments where the lattice friction on screw dislocations in Zr-O alloys 
has been attributed to the presence of jogs on the dislocations due to local cross-slip.
\end{abstract}


\end{frontmatter}

\section{Introduction}

Zirconium is a hexagonal close-packed (hcp) transition metal, which deforms mainly by glide of $\hkl<a>=1/3\,\hkl<1-210>$ dislocations in \hkl(10-10) prismatic planes \cite{Rapperport1959,Rapperport1960,Caillard2003}. All experimental studies \cite{Tyson1967a,DasGupta1968,Levine1968,Mills1968,Baldwin1968,Soo1968,Akhtar1971,Sastry1971,Akhtar1975b,Akhtar1975c,Caillard2003} point to a low-temperature behavior controlled by lattice friction acting on screw dislocations. This lattice friction is extrinsic in nature and arises due to interactions with interstitial solute atoms like oxygen, as indicated by the rapid increase of the critical resolved shear stress for prismatic slip in single crystals when the oxygen content increases \cite{Mills1968,Baldwin1968,Soo1968,Akhtar1971}. Moreover, microscopic observations of deformed Zr alloys \cite{Bailey1962,Akhtar1971,Ferrer2002} show long rectilinear screw dislocations, whose mobility is therefore limited compared to other orientations.

By way of contrast, in pure zirconium, dislocations of all characters have similar mobilities \cite{Clouet2015} and the plastic behavior becomes almost athermal like in fcc metals for very low O contents ($\lesssim100$\,ppm in weight) \cite{Mills1968,Soo1968}. The easy prismatic glide of screw dislocations in pure Zr is consistent with \abinitio calculations, which show that screw dislocations have a ground state dissociated in a prismatic plane \cite{Ferrer2002,Domain2000,Clouet2012,Clouet2015}, with a small energy barrier opposing prismatic glide and a corresponding Peierls stress below 21\,MPa \cite{Clouet2012}. Screw dislocations in Zr can adopt other configurations with a core partly or totally  spread in a first-order pyramidal plane \cite{Chaari2014,Chaari2014a,Clouet2015}. These configurations however are of higher energy and contribute only to secondary slip, which is activated above room temperature.

Oxygen is therefore added to zirconium alloys for its strengthening effect at an average content between 600 and 1200 wppm Zr (0.3-0.7 at.\%) and with a maximum accepted content of 2000 wppm (1.2 at.\%) \cite{Lemaignan2012}.
The same hardening of prismatic slip by O solutes has been reported in titanium \cite{Churchman1954,Tyson1967a,Akhtar1975a,Conrad1981,Naka1988,Biget1989,Caillard2003}.
However, unlike in Zr, an intrinsic lattice friction exists even in pure Ti, because in this metal, the screw dislocations glide by a locking-unlocking mechanism due to the fact that their ground state is dissociated in a first-order pyramidal plane~\cite{Clouet2015}.

The origin of hardening caused by an O addition in Zr and Ti is not clearly understood. 
A control of dislocation glide through their interaction with interstitial solute atoms,
as described by Fleisher model \cite{Fleischer1962}, has been proposed \cite{Levine1968,Tanaka1972,Akhtar1975a}, 
but it is not clear why such interaction would mainly affect the screw dislocations.
Moreover, this interaction must involve dislocation cores \cite{Tiwari1972,Naka1988}, 
since no elastic interaction exists in an hcp crystal between an \hkl<a> screw dislocation and an interstitial solute atom in an octahedral interstitial site \cite{Tyson1967a}.
Other authors attributed the lattice friction
to the presence of jogs on the screw dislocations, 
with an increase of the jog density with the O content \cite{Bailey1962,Mills1968,Soo1968,Akhtar1975b}.
\Abinitio calculations  in Ti \cite{Yu2015} have evidenced a short range repulsive interaction
between screw dislocations and oxygen atoms, 
which can lead to solid solution hardening and to the creation of jogs by local dislocation cross-slip.

The aim of the present article is to characterize the short-range interactions between \hkl<a> screw dislocations and O atoms in Zr in order to better understand the effect of O addition on the lattice friction in this metal. Since core effects are expected, we will use an electronic structure description of atomic interactions based on \abinitio calculations. We first analyse the interaction of O atoms with the prismatic and pyramidal stacking faults involved in the various potential dissociations of the screw dislocation. Interaction of an oxygen atom with the three different stable and metastable configurations of an \hkl<a> screw dislocation is then investigated. Consequences of the repulsive interaction evidenced by these calculations are discussed in the last section.

\section{Methods}
\label{sec:methods}

\Abinitio calculations are performed with the {\pwscf} code \cite{Giannozzi2009}.
We use the generalized gradient approximation with the exchange-correlation functional of Perdew, Burke and Ernzerhof \cite{Perdew1996}. Zr and O atoms are modelled with an ultrasoft Vanderbilt pseudo-potential, with 4s and 4p semi-core electrons included in the valence states of Zr. Electronic wave functions are described with plane waves with an energy cutoff of 28\,Ry. A regular grid is used for the integration in reciprocal space, with $14\times14\times8$ k-points for the primitive hcp cell and an equivalent k-point density for larger supercells. The electronic density of state is broadened with the Methfessel-Paxton function, with a broadening of 0.3\,eV. Atoms are relaxed until all atomic force components are below 10\,meV/\AA. The same set of \abinitio parameters has already been shown to predict Zr lattice parameters and elastic constants in good agreement with experimental data \cite{Clouet2012} and to provide an accurate description of \hkl<a> screw dislocations in pure zirconium \cite{Clouet2012,Chaari2014,Clouet2015}.

\begin{table}[!tb]
	\caption{Energy of an O atom in a perfect hcp lattice for different configurations:
	octaheadral (O) taken as the reference, tetrahedral (T), basal tetrahedral (BT) 
	and crowdion (C). 
	Results are given for different supercells defined by their number $N_{\rm sites}$ of lattice sites.
	The supercell is based either on the hcp primitive cell 
	or corresponds to the cell used for dislocation modeling
	(\cf \S \ref{sec:dislo_methods}).}
	\label{tab:energy_O}
	\centering
	\begin{tabular}{cccccc}
		\hline
		\multirow{2}{*}{cell}	& \multirow{2}{*}{$N_{\rm sites}$}	& \multicolumn{4}{c}{$\Delta E$ (eV)} \\
		\cline{3-6}
					&	& O	& T	& BT	& C \\
		\hline
		$3\times3\times2$ hcp	& 36	& 0.	& 0.90 & 0.93	& 1.82	\\
		$4\times4\times3$ hcp	& 96	& 0.	& 0.87 & 0.91	& 1.90	\\
		$5\times5\times4$ hcp	& 200	& 0.	& 0.84 & 0.87	& 1.91	\\
		$8\times5\times2$ dislo	& 320	& 0.	& 0.90 & 0.95	& 	\\
		%
		\hline
	\end{tabular}
\end{table}

We checked that the most stable position for O atoms
in the hcp Zr lattice is the octahedral interstitial site.  
Other interstitial positions -- tetrahedral, basal tetrahedral and crowdion sites --
have a higher energy (Tab. \ref{tab:energy_O}). 
The energy difference between these various interstitial configurations varies only slightly with the size of the supercell, showing that there is no complex long-range interaction between O atoms and their periodic images.

\section{Oxygen interaction with stacking faults}

\hkl<a> screw dislocations in Zr can adopt several configurations with either a planar or a non-planar dissociation in either a prismatic or a pyramidal-I plane, or a combination of both \cite{Clouet2015}. We therefore study the interaction of O atoms with the corresponding prismatic and pyramidal stacking faults
\cite{Clouet2012,Chaari2014,Chaari2014a}, as a first step towards modeling the interaction of O atoms with screw dislocations.

\subsection{Simulation setup}

The interaction energy of an O atom with a stacking fault is defined as
\begin{equation*}
	E^{\rm int}= E_{\textrm{SF-O}} - E_{\rm O} - E_{\rm SF} + E_{\rm bulk},
\end{equation*}
where $E_{\textrm{SF-O}}$, $E_{\rm O}$, $E_{\rm SF}$, and $E_{\rm bulk}$ are the energies
respectively of the cell containing both the O atom and the stacking fault,
the cell with only the O atom,
the cell with only the stacking fault,
and the perfect hcp bulk cell.
The same tri-periodic simulation cell is used for all four calculations, 
with only a shift equal to the fault vector added to the periodicity vector perpendicular to the fault plane
for $E_{\rm SF-O}$ and $E_{\rm SF}$ \cite{Clouet2012}.
Only one fault and one O atom are introduced in the simulation cells.
The O atom is placed in the already relaxed stacking fault
and full atomic relaxations, without constraint, are then performed. 

The lengths of the periodicity vectors are defined by three integers, $n$, $m$, and $p$. 
For the prismatic stacking fault, these vectors are 
$n/3\,\hkl[1-210]$, $m\,\hkl[0001]$, and $p\,\hkl[10-10]$,
and for the pyramidal fault $n/3\,\hkl[1-210]$, $m/3\,\hkl[2-1-13]$, and $p/3\,\hkl[-1-123]$.
In both cases, the number of atoms in the simulation cell is $4nmp$.
The integers $n$ and $m$ fix the surface of the stacking fault,
and thus the fraction of solute atom per fault area,
whereas $p$ controls the distance between the fault plane and its periodic images.
We have checked the convergence of the calculations with respect to these cell dimensions in  
Tables \ref{tab:fault_pr} and \ref{tab:fault_py}.

\subsection{Prismatic stacking fault}
\label{sec:fault_prism}

\begin{figure}[!b]
	\begin{center}
		\subfigure[\hkl{10-10} fault]{\includegraphics[scale=0.36]{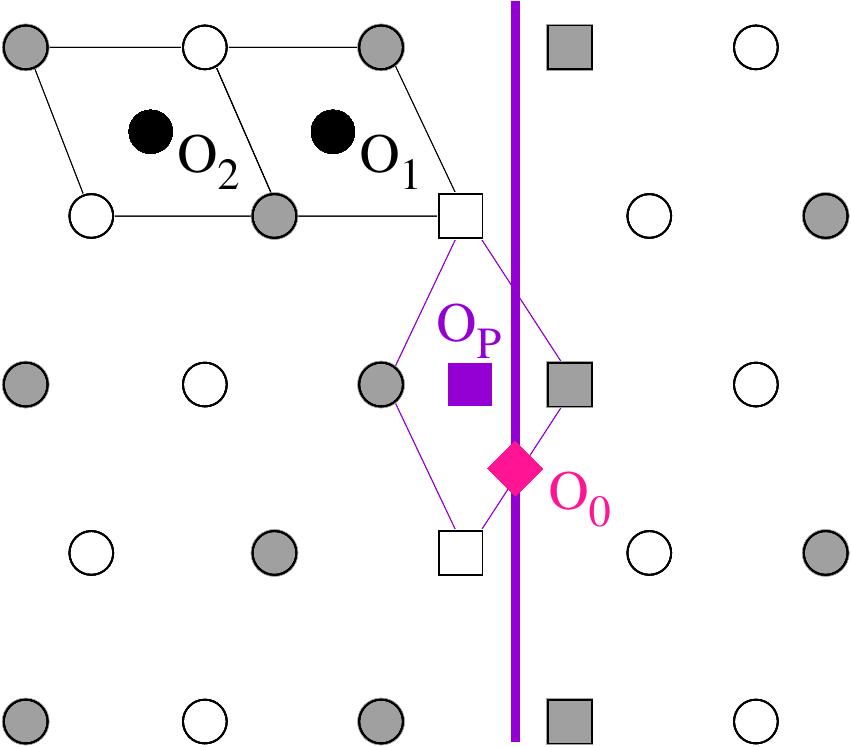}}
		\hfill
		\subfigure[\hkl{10-1-1} fault]{\includegraphics[scale=0.36]{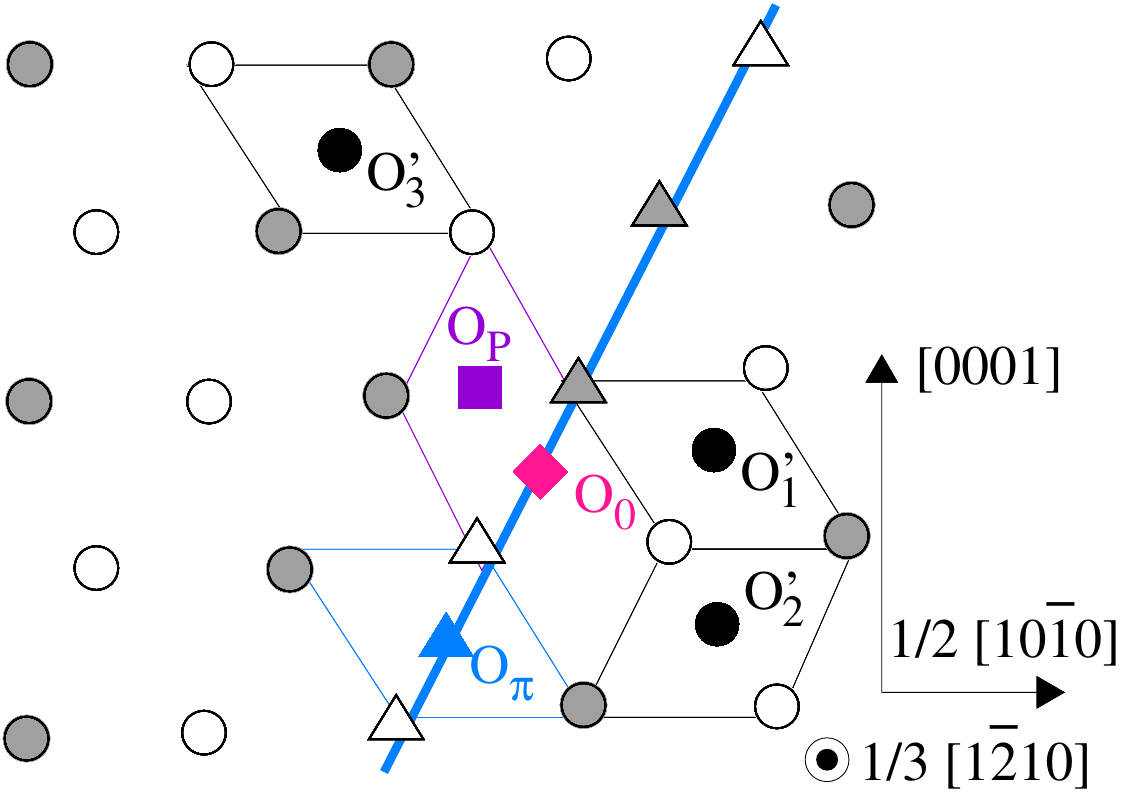}}
	\end{center}
	\caption{Lattice positions of an oxygen interstitial in the vicinity 
	of (a) a prismatic and (b) a pyramidal stacking fault, shown in a $\hkl[1-210]$ projection.
	$O_1$, $O_2$, $O'_1$, $O'_2$, and $O'_3$ are octahedral sites slightly distorted by the fault
	(circle symbols), 
	$O_0$ is an octahedral site destroyed by the fault (diamond symbol),
	and $O_{\rm P}$ and $O_{\pi}$ are new octahedral sites created by the fault
	(respectively square and triangle symbols).
	Zr atoms are shown with grey and white symbols depending on their position 
	along the $\hkl[1-210]$ direction in the faulted crystal.}
	\label{fig:fault}
\end{figure}

\begin{table}[!b]
	\caption{Interaction energies (meV) of an oxygen atom with a prismatic stacking fault
	for different positions of the oxygen interstitial (see Fig. \ref{fig:fault}a).
	Calculations are performed for different $n \times m \times p$ cell sizes 
	corresponding to the periodicity vectors
	$n/3\,\hkl[1-210]$, $m\,\hkl[0001]$, and $p\,\hkl[10-10]$.}
	\label{tab:fault_pr}
	\begin{center}
	\begin{tabular}{cccccc}
		 \hline  
		 site  		& 4$\times$4$\times$3 & 4$\times$2$\times$4 & 2$\times$4$\times$4	& 4$\times$4$\times$4 & 5$\times$5$\times$4 \\
		 \hline
		 $O_{\rm P}$  	&  164 &  188	& 166	& 161	& -        \\
		 $O_1$  	&  33  &   -  	& -  	& 26 	&  27    \\
		 $O_2$  	&  44  &   -  	&  - 	& 40 	&  40   \\
		 \hline 
	\end{tabular} 
	\end{center}
\end{table}

The octahedral sites located in the fault plane (sites $O_0$ in Fig. \ref{fig:fault}a) are destroyed by the shear associated with the stacking fault. These sites become unstable positions for the O atom, which migrates during atomic relaxation to a nearby new octahedral site created by the fault. This site, labelled $O_{\rm P}$ in Fig. \ref{fig:fault}a, has a geometry close to an octahedral site in the body centered cubic (bcc) structure \cite{Ghazisaeidi2014a}. It is a stable position for the O atom, without any reconstruction of the fault, but the interaction energy is positive, and therefore repulsive (Tab. \ref{tab:fault_pr}). This is in agreement with oxygen being an $\alpha$-stabilizer since O is then expected to have a higher energy in the bcc than in the hcp structure. A similar result has been reported in Ti \cite{Ghazisaeidi2014a,Yu2015,Kwasniak2016}.

A slightly repulsive energy is also obtained for the octahedral sites located in the immediate vicinity of the fault plane (sites $O_1$ and $O_2$ in Fig. \ref{fig:fault}a) (see Tab. \ref{tab:fault_pr}). There is therefore no attractive site and oxygen atoms can only increase the prismatic stacking fault energy in Zr.

\subsection{First-order pyramidal stacking fault}

\begin{table}
	\caption{Interaction energies (meV) of an oxygen atom with a pyramidal stacking fault
	for different positions of the oxygen interstitial (see Fig. \ref{fig:fault}b).
	Calculations are performed for different $n \times m \times p$ cell sizes 
	corresponding to the periodicity vectors
	$n/3\,\hkl[1-210]$, $m/3\,\hkl[2-1-13]$, and $p/3\,\hkl[-1-123]$.}
	\label{tab:fault_py}
	\begin{center}
	\begin{tabular}{ccccc}
		\hline 
		site 		& 4$\times$2$\times$4 & 4$\times$3$\times$4 & 6$\times$3$\times$4 & 4$\times$3$\times$5  \\
		\hline
		$O_{\rm P}$	& 130   & 129	& 131	& 123       \\
		$O_{\pi}$	&  30   &  22 	& 29   	&  -  \\
		$O'_1$		&  75   &  72 	& 77 	&  34 \\
		$O'_2$		&   0   &   0  	&  0   	&  10  \\
		$O'_3$		&   0   &   0 	&  0 	&   0   \\
		\hline 
	\end{tabular}
	\end{center}
\end{table}

The pyramidal fault also destroys 
the octahedral sites located in the fault plane (sites $O_0$ in Fig. \ref{fig:fault}b)
and creates new bcc-like octahedral sites (sites $O_{\rm P}$ in Fig. \ref{fig:fault}b) that are stable insertion sites but with a 
repulsive interaction energy (Tab. \ref{tab:fault_py}), although slightly less than with the prismatic fault (Tab. \ref{tab:fault_pr}).

Since the pyramidal stacking fault corresponds to a two-layer twin in the pyramidal stacking \cite{Chaari2014,Chaari2014a}, new octahedral sites, labelled $O_{\pi}$ in Fig. \ref{fig:fault}b, are created and correspond to octahedral sites in the twinned hcp crystal. These are stable insertion sites for O atom, although still with a slightly repulsive interaction energy (Tab. \ref{tab:fault_pr}).

Finally, a repulsive interaction energy is also obtained for the octahedral interstitial sites in the close vicinity of the fault (sites labelled $O'_1$, $O'_2$, and $O'_3$ in Fig. \ref{fig:fault}b), but this interaction energy rapidly vanishes away from the fault (Tab. \ref{tab:fault_py}).

In conclusion, no attractive insertion site exists for O atoms inside or in the vicinity of neither a prismatic nor a pyramidal-I stacking fault in Zr.

\section{Oxygen interaction with dislocations}

\subsection{Simulation setup}
\label{sec:dislo_methods}

Dislocations are modeled in simulation cells with full periodic boundary conditions, which requires to introduce a dipole of dislocations of opposite Burgers vectors \cite{Rodney2017}. A quadrupolar periodic array of dislocations, described as an S arrangement in Ref. \cite{Clouet2012} is used. The periodicity vectors of the simulation cell, before introduction of the dislocation dipole, are
$\vec{a}_x = n \,a[10\bar{1}0]$, 
$\vec{a}_y = m \,c[0001]$
and $\vec{a}_z = p\,\vec{b} = p/3 \ a [1\bar{2}10]$,
where $n$, $m$ and $p$ are three integers.
Most of the calculations were performed 
with $n=5$, $m=8$ and $p=2$, corresponding to a simulation cell with 320 Zr atoms.
A $2\times1\times7$ k-point grid was used for this simulation cell. 

\begin{figure}[!b]
	\begin{center}
		\subfigure[prismatic]{\includegraphics[width=0.37\linewidth]{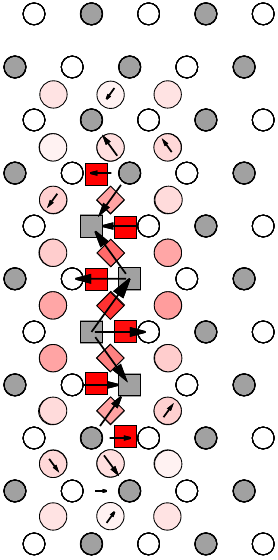}}
		\hfill
		\subfigure[pyramidal]{\includegraphics[width=0.37\linewidth]{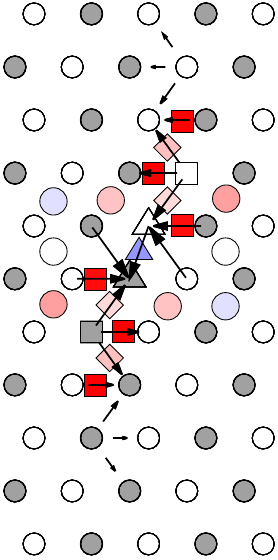}}
		\includegraphics[width=0.16\linewidth]{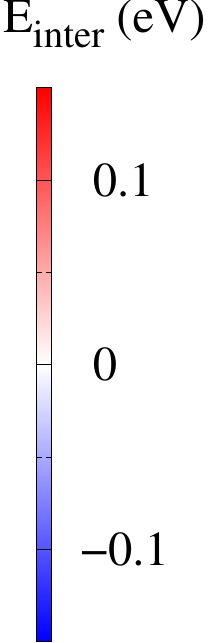}
	\end{center}
	\caption{Interaction energy between an oxygen atom and a screw dislocation when the dislocation is (a) fully dissociated in a prismatic plane and (b) partly dissociated in a pyramidal plane and in two prismatic planes.
	The differential displacement maps show the dislocation core structure
	before introduction of the oxygen atom. 
	The colored symbols correspond to the different positions where a single oxygen atom 
	has been inserted, with a color depending on the interaction energy.
	Different symbols are used for different insertion sites (\cf Fig. \ref{fig:fault}).}
	\label{fig:dislo_O_inter}
\end{figure}

\begin{figure*}[!bth]
	\centering
	\begin{tabular}{c|c|cccc}
		& Initial  \\
		& configuration & \multicolumn{4}{c}{\raisebox{1.5ex}[0cm][0cm]{Relaxed configurations} }\\
		\hline
		\rotatebox{90}{\parbox{0.29\linewidth}{\centering Sites destroyed \\ by the stacking fault}}
		&
		a)\includegraphics[width=0.14\linewidth]{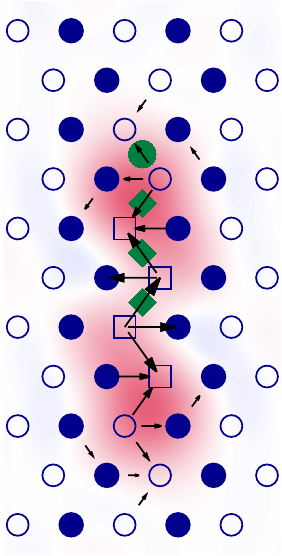}
		&
		b)\includegraphics[width=0.14\linewidth]{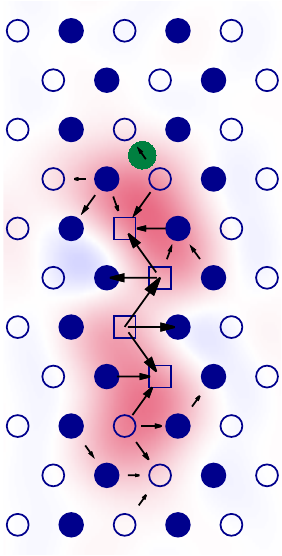}
		&
		c)\includegraphics[width=0.14\linewidth]{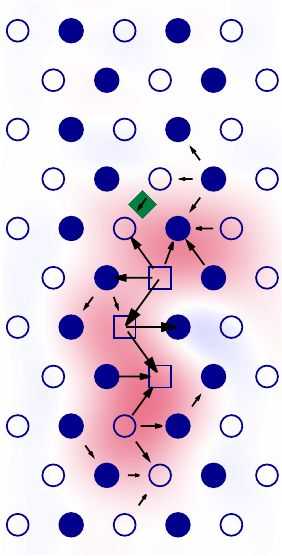}
		&
		d)\includegraphics[width=0.14\linewidth]{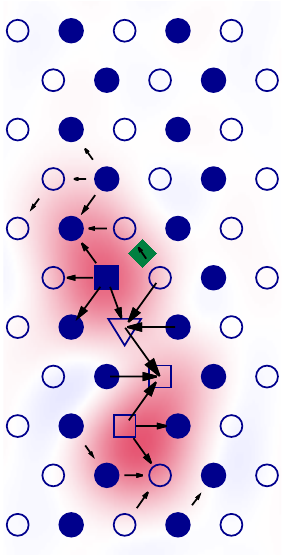}
		&
		e)\includegraphics[width=0.14\linewidth]{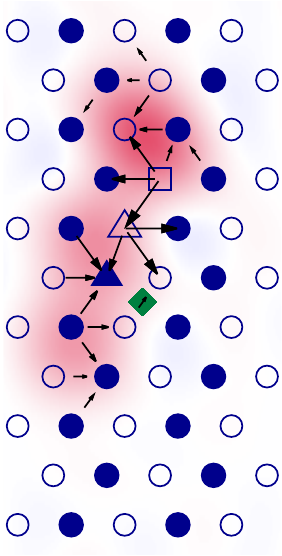}
		\\
		\hline
		\rotatebox{90}{\parbox{0.29\linewidth}{\centering Sites created \\ by the stacking fault}}
		&
		f)\includegraphics[width=0.14\linewidth]{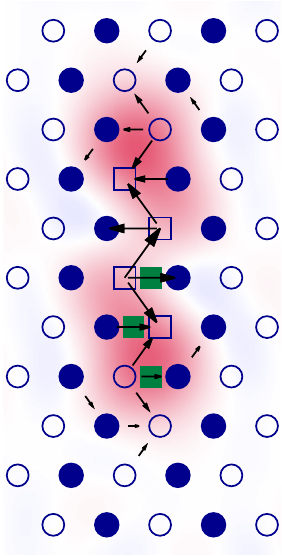}
		&
		g)\includegraphics[width=0.14\linewidth]{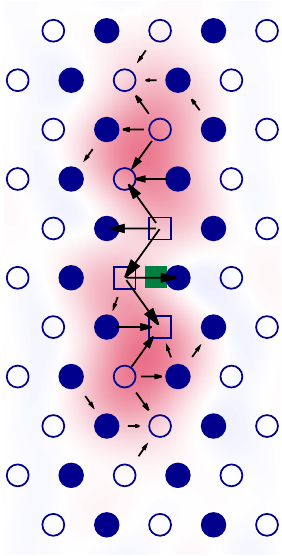}
		&
		h)\includegraphics[width=0.14\linewidth]{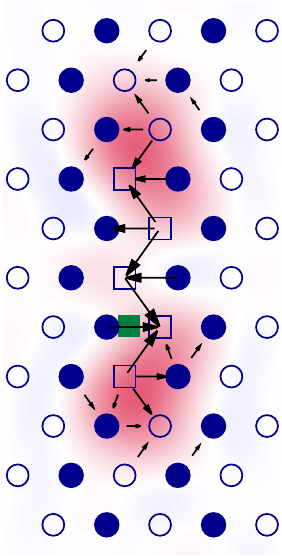}
		&
		i)\includegraphics[width=0.14\linewidth]{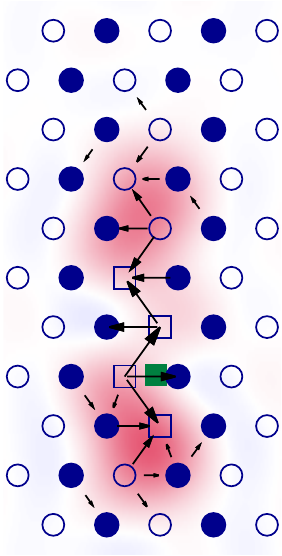}
	\end{tabular}
	\caption{Core structure of a screw dislocation initially dissociated in a prismatic plane 
	in presence of an O impurity. 
	The initial configuration of the dislocation, before introduction of the O atom, is shown in (a) and (f)
	with the various insertion sites for the O atom.
	The relaxed structures are shown in (b-e) for an O atom lying in an octahedral site 
	destroyed by the prismatic spreading of the dislocation
	and in (g-i) for an O atom in an octahedral site created by the prismatic fault.
	The position of the O atom is shown with a green symbol 
	(see Fig. \ref{fig:fault} for a definition of the different symbols).
	The contour map shows the dislocation density according to the Nye tensor.}
	\label{fig:dislo_O_prism}
\end{figure*}

\begin{figure*}[!bth]
	\centering
	\begin{tabular}{c|c|cccc}
		& Initial  \\
		& configuration & \multicolumn{4}{c}{\raisebox{1.5ex}[0cm][0cm]{Relaxed configurations} }\\
		\hline
		\rotatebox{90}{\parbox{0.29\linewidth}{\centering Sites destroyed \\ by the stacking fault}}
		&
		a)\includegraphics[width=0.14\linewidth]{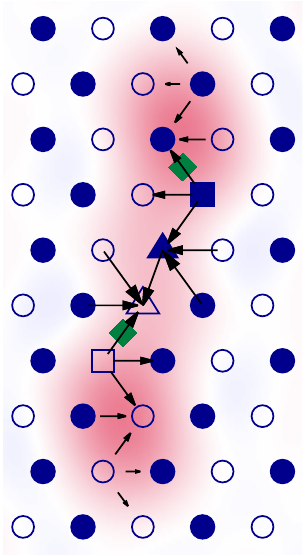}
		&
		\multicolumn{2}{c}{b)\includegraphics[width=0.14\linewidth]{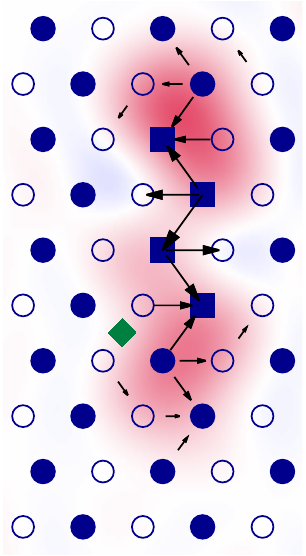}}
		&
		\multicolumn{2}{c}{c)\includegraphics[width=0.14\linewidth]{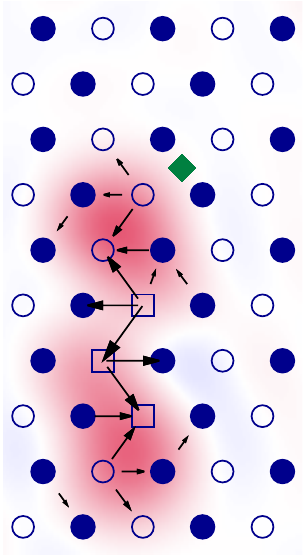}}
		\\
		\hline
		\rotatebox{90}{\parbox{0.29\linewidth}{\centering Sites created \\ by the stacking fault}}
		&
		d)\includegraphics[width=0.14\linewidth]{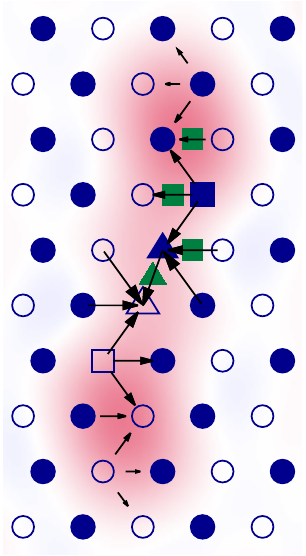}
		&
		e)\includegraphics[width=0.14\linewidth]{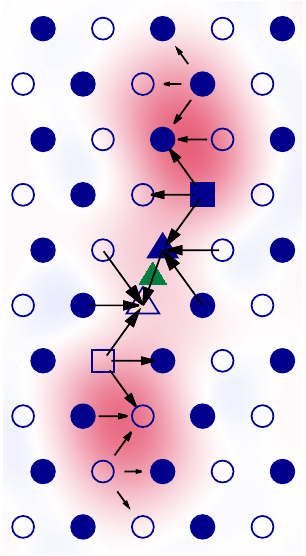}
		&
		f)\includegraphics[width=0.14\linewidth]{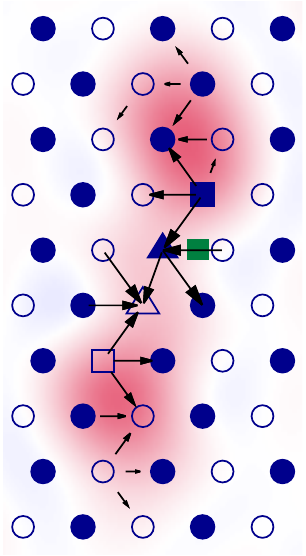}
		&
		g)\includegraphics[width=0.14\linewidth]{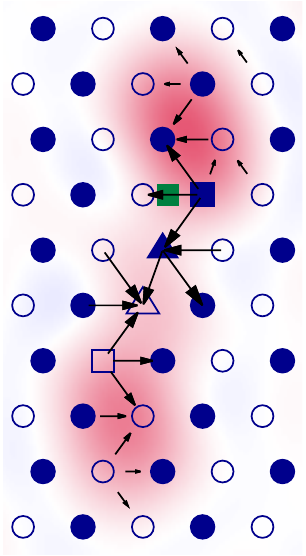}
		&
		h)\includegraphics[width=0.14\linewidth]{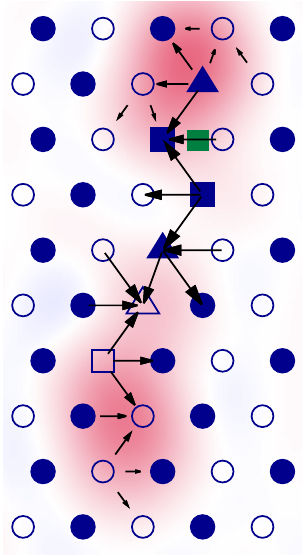}
	\end{tabular}
	\caption{Core structure of a screw dislocation initially partly spread in a pyramidal plane and partly in two adjacent prismatic planes, in presence of an O impurity. 
	The initial configuration of the dislocation, before introduction of the O atom, is shown in (a) and (d)
	with the various insertion sites for the O atom.
	The relaxed structures are shown in (b-c) for octahedral sites 
	destroyed by the prismatic spreading of the dislocation,
	in (e) for the octahedral site created by the pyramidal fault
	and in (f-h) for octahedral sites created by the prismatic faults.
	}
	\label{fig:dislo_O_py}
\end{figure*}

We considered the three different configurations that a screw dislocation can adopt in Zr \cite{Clouet2015}:
the ground state dissociated in a prismatic plane 
(Figs. \ref{fig:dislo_O_inter}a and \ref{fig:dislo_O_prism}) and two metastable configurations with either a nonplanar core spread in a combination of prismatic and pyramidal planes~ \cite{Chaari2014,Clouet2015}
(Figs. \ref{fig:dislo_O_inter}b and \ref{fig:dislo_O_py}) or a planar core spread 
in a pyramidal plane \cite{Clouet2015} (Fig. \ref{fig:dislo_O_py_inter}).
In a $5\times8\times p$ simulation cell, the relative energies of the metastable configurations
are $\Delta E_1 = 1.7$ and $\Delta E_2 = 12.1$\,meV\,\AA$^{-1}$, respectively.

One oxygen atom is introduced in the simulation cell in an interstitial site close to one of the dislocations of the dipole. Introducing two oxygen atoms in equivalent positions near both dislocations is also obviously possible but leads to slower convergence when the dislocation cores reconstruct during relaxation. Moreover, in section \ref{sec:O_segregation}, we will check that introducing one or two solutes leads to similar energies. Also, most of the calculations were performed with a dislocation length $l_z=2b$ with $p=2$. There is therefore one O atom every other repeating distance along the dislocation line. This corresponds to a very high concentration of O atoms, but in section \ref{sec:O_segregation}, we will also check that the O-dislocation interaction energy does not depend on the distance between O atoms along the dislocation line as long as $p\ge2$.

The interaction energy is defined as 
\begin{equation}
	E^{\rm int} = E_{\rm Dislo-O} -  E_{\rm O} -  E_{\rm Dislo} + E_{\rm bulk},
	\label{eq:Einter}
\end{equation}
where $E_{\rm Dislo-O}$, $E_{\rm O}$, $E_{\rm Dislo}$, and $E_{\rm bulk}$
are the energies of the same $5\times8\times2$ cell 
containing respectively both the dislocation and the O atom,
only the O atom,
only the dislocation,
and no defect nor solute.
The reference energy for the dislocation, $E_{\rm Dislo}$, 
is the energy of the initial dislocation configuration before introduction of the O atom.
This definition of the interaction energy assumes that the elastic energy 
of the dislocation periodic array is the same with or without the O atom, 
which is a valid approximation when no core reconstruction is induced by the solute (\cf \S \ref{sec:O_segregation}).

The relaxed dislocation core structures are visualized using both differential displacements and dislocation density contour maps  in Figs. \ref{fig:dislo_O_prism}, \ref{fig:dislo_O_py} and \ref{fig:dislo_O_py_inter}. 
In these projections in the plane perpendicular to the dislocation line,
Zr atoms are sketched by empty and filled symbols depending on their \hkl(1-210) plane in the initial perfect crystal.
Different symbols are used depending on the neighbourhood of the Zr atom.
Differential displacement maps \cite{Vitek1970} show the strain created by the screw dislocation.
In these maps, arrows between neighbouring \hkl[1-210] atomic columns have an amplitude proportional to the \hkl[1-210] component of the differential displacement between the two columns.  Displacements smaller than $0.1b$ are not shown for clarity.
Contour maps of the dislocation density, deduced from the screw component
along the \hkl[1-210] direction of the Nye tensor, are also shown.
The Nye tensor is extracted from the relaxed atomic structures
using the method developed by Hartley and Mishin \cite{Hartley2005}.
For post-processing, the neighbourhood of an atom is defined 
by a sphere of radius 1.15 times the lattice parameter.
This neighbourhood is identified with one of the following perfect patterns:
same neighbourhood as in a perfect hcp lattice (circles),
in an unrelaxed prismatic stacking fault (squares), 
and in a perfect hcp lattice after a \hkl(-1011) or a \hkl(10-11) mirror symmetry to correspond with the two variants of the unrelaxed pyramidal stacking fault \cite{Chaari2014,Chaari2014a} (upward and downward triangles). 
The angle threshold on atomic bonds was set to 10$^{\circ}$.

\subsection{Dislocation prismatic configuration}

When the dislocation is in its ground state, \ie dissociated in a prismatic plane, 
a repulsive interaction with the O atom is obtained for all potential insertion sites of the solute
(Fig. \ref{fig:dislo_O_inter}a).  The interaction is only slightly repulsive 
when the O atom sits in an octahedral site away from the dislocation habit plane. 
On the other hand, when the octahedral site belongs to the prismatic plane of dissociation, 
the interaction increases up to 0.12\,eV as the oxygen gets closer to the dislocation center.
This repulsive interaction in the dislocation core is due to the prismatic stacking fault
created by the dislocation dissociation, which destroys the octahedral  
sites of the hcp lattice, as discussed in section \ref{sec:fault_prism} and shown as green diamonds in Fig. \ref{fig:dislo_O_prism}a-e.  
During relaxation, the O atom remains immobile but as seen in Fig. \ref{fig:dislo_O_prism}c-e, the screw dislocation cross-slips to restore the initial octahedral site around the O atom. 
Such a cross-slip corresponds to a conservative motion of one half of the stacking fault perpendicular to itself, as already evidenced in pure Zr \cite{Chaari2014}.
The relaxed dislocation is then partly spread in the initial prismatic plane and partly in a cross-slipped pyramidal plane. 

The prismatic stacking fault also creates new octahedral insertion sites 
(sites labelled $O_{\rm P}$ in Fig. 1a), which can also be found in the dislocation core (green squares in Fig. \ref{fig:dislo_O_prism}f-i). Upon relaxation, they all lead to almost identical configurations, with repulsive energies between 0.15 and 0.20\,eV (Fig. \ref{fig:dislo_O_inter}a), close to the value obtained with a perfect stacking fault (Tab. \ref{tab:fault_pr}). In all cases, the dislocation remains fully dissociated in its prismatic habit plane and glides during relaxation such that the O atom is positioned near the inner side of the closest Shockley partial. One can only see a slight variation of the dislocation prismatic spreading in Figs. \ref{fig:dislo_O_prism}g-i. 

\subsection{Dislocation pyramidal configurations}

We now examine the interaction of an O atom with the pyramidal configurations of the screw dislocation, starting with the configuration of lower energy, which exhibits a non-planar dissociation in a pyramidal plane and two adjacent prismatic planes \cite{Chaari2014}. When the O atom sits in an octahedral site outside the stacking faults, the interaction is small, between -0.02 and 0.05\,eV (Fig. \ref{fig:dislo_O_inter}b). Contrary to the prismatic configuration, at least one attractive position exists, located right above the pyramidal stacking fault created by the dislocation dissociation. This slight attraction ($-0.02$\,eV) may result from the elastic interaction of the O atom with the two edge disconnections, which connect the pyramidal stacking fault to the two prismatic faults \cite{Chaari2014}.

When the O atom lies in the prismatic parts of the stacking fault ribbon, the same repulsive interaction is observed as with the prismatic configuration. For octahedral sites destroyed by the prismatic stacking fault, the O atom repels the dislocation, which cross-slips back in its prismatic ground state (Fig. \ref{fig:dislo_O_py}b-c). For octahedral sites created by the prismatic stacking fault (Fig. \ref{fig:dislo_O_py}f-h), no core reconstruction nor solute migration occurs and the interaction is repulsive (between 0.14 and 0.26\,eV).

The octahedral site created by the pyramidal stacking fault (sites $O_{\pi}$ in Fig. \ref{fig:fault}b) is also found in the dislocation core (upward triangle in Fig. \ref{fig:dislo_O_inter}b and Fig. \ref{fig:dislo_O_py}d,e). This site, located exactly at the dislocation center, leads to an attractive interaction ($-0.06$\,eV), without modification of the dislocation configuration apart from a slight increase of the spreading (Fig. \ref{fig:dislo_O_py}e). 
The interaction of the O atom in this $O_{\pi}$ site with the perfect stacking fault was found slightly repulsive (Tab. \ref{tab:fault_py}). 
This example shows that although calculations with stacking faults are a reasonable surrogate to rationalize solute interaction with dislocation cores,
they cannot fully substitute a real atomic modeling of the dislocation core

\begin{figure}[!t]
	\centering
	\begin{tabular}{c|cc}
		 Initial  \\
		 configuration & \multicolumn{2}{c}{\raisebox{1.5ex}[0cm][0cm]{Relaxed configurations} }\\
		 \hline
		\includegraphics[width=0.28\linewidth]{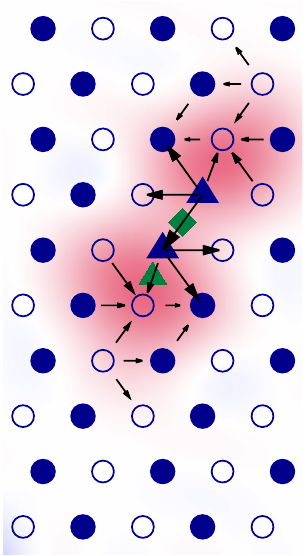}
		&
		\includegraphics[width=0.28\linewidth]{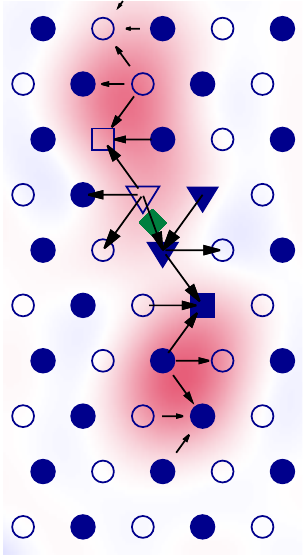}
		&
		\includegraphics[width=0.28\linewidth]{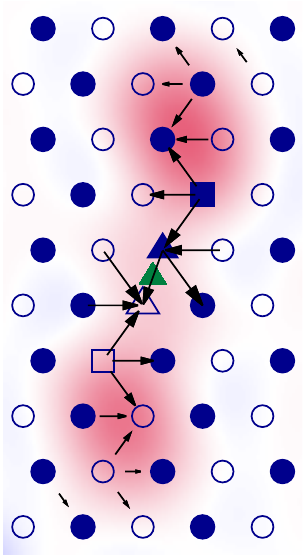} 
		\\
		\multicolumn{1}{l}{a)} & \multicolumn{1}{l}{b)} & \multicolumn{1}{l}{c)}
	\end{tabular}
	\caption{Core structure of a screw dislocation initially dissociated in the pyramidal plane 
	in presence of an O impurity. 
	The initial configuration of the dislocation, before introduction of the O atom, is shown in (a)
	with the various insertion sites.
	The relaxed structures are shown in (b) for an octahedral site 
	destroyed by the pyramidal spreading of the dislocation and
	in (c) for an octahedral site created by the pyramidal fault.
	}
	\label{fig:dislo_O_py_inter}
\end{figure}

The same attractive configuration is also obtained when an O atom interacts
with the metastable core of higher energy, which is planar and fully dissociated in a pyramidal plane
(Fig. \ref{fig:dislo_O_py_inter}).
We only considered insertion sites in the stacking fault ribbon arising from the pyramidal dissociation.
The position corresponding to the dislocation center is an octahedral site destroyed by the pyramidal stacking fault (green diamond in Fig. \ref{fig:dislo_O_py_inter}a,b).
When an O atom is inserted at this site, the dislocation relaxes to its metastable nonplanar configuration 
of lower energy and the O atom shuffles to the attractive position seen above (Fig. \ref{fig:dislo_O_py_inter}b).
If inserted directly in this new octahedral site, the O atom
does not move but the dislocation glides in its pyramidal habit plane to position its center at
to the O position (Fig. \ref{fig:dislo_O_py_inter}c) and the dislocation core again reconstructs to adopt the metastable nonplanar configuration of lower energy.

\subsection{Oxygen segregation}
\label{sec:O_segregation}

\begin{figure}[!b]
	\begin{center}
		\includegraphics[width=0.8\linewidth]{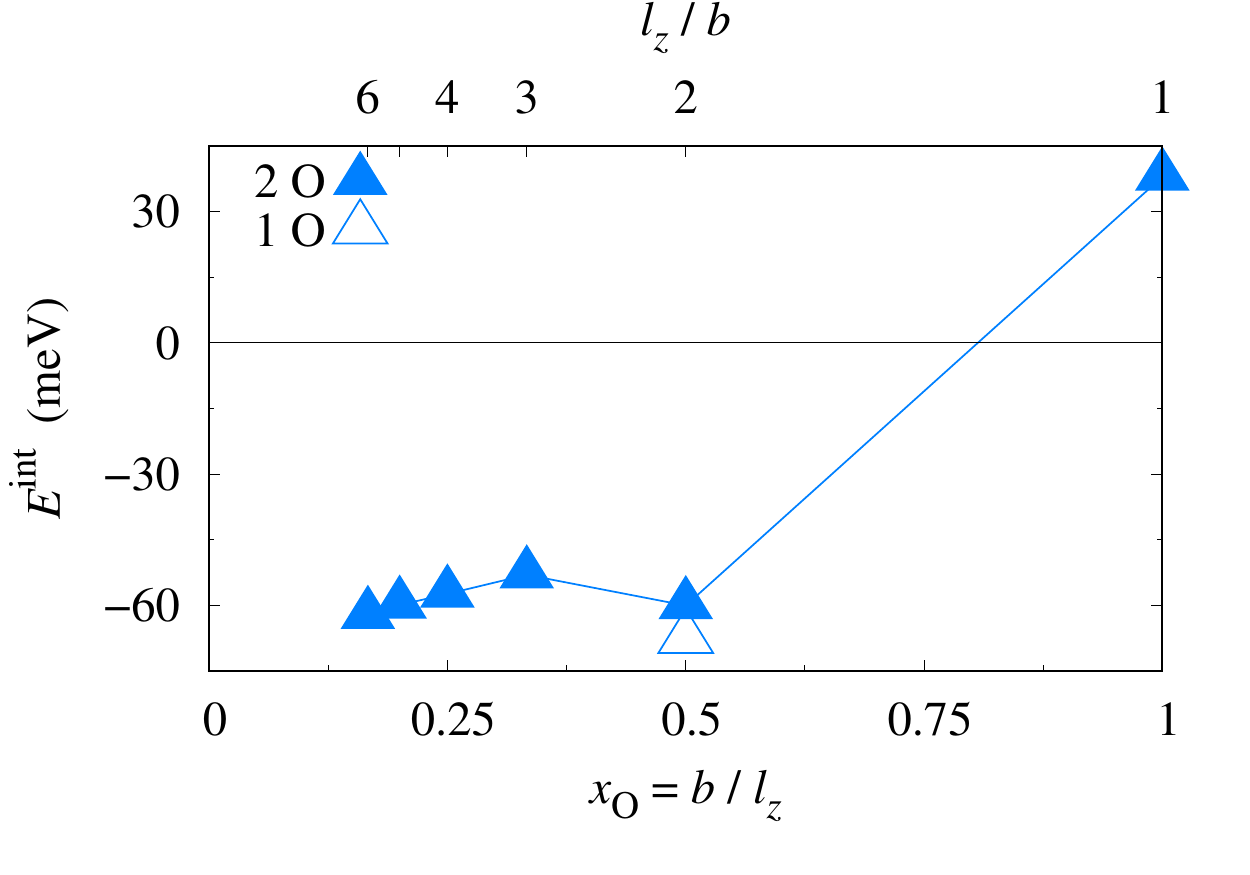}
	\end{center}
	\caption{Interaction energy $E^{\rm int}$ of an oxygen atom with a screw dislocation 
	when the oxygen is inserted at the center of the pyramidal configuration
	(site $O_{\pi}$, Fig. \ref{fig:dislo_O_py}e)
	as a function of the atomic fraction $x_{\rm O} = b/l_z$ of solute atoms on the dislocation lines. 
	Calculations were performed using either two solute atoms, one on each dislocation of the dipole, 
	or a single solute on one of the dislocations.}
	\label{fig:Eint_O_py}
\end{figure}

In previous subsection, we showed that when the dislocation is in its nonplanar metastable configuration,
the insertion site $O_{\pi}$ at its center (Fig. \ref{fig:dislo_O_py}e) is attractive
and can therefore lead to segregation. 
We now look at how this attractive interaction varies for various oxygen contents along the dislocation line, using $5\times8\times p$ simulation cells with different dislocation lengths $l_z = p\,b$. As only one insertion site $O_{\pi}$ exists per periodicity length $b$, the atomic fraction of oxygen segregated on the line is $x_{\rm O} = b / l_z = 1/p$. Calculations were performed inserting either one solute atom on one of the dislocations of the dipole, or one solute on each dislocation of the dipole. 
For the atomic fraction $x_{\rm O}=1/2$ ($5\times8\times2$ supercell), using one or two O atoms leads 
only to a small difference (8\,meV) on the interaction energy (Fig. \ref{fig:Eint_O_py}).
This shows that, in absence of core reconstruction, the variation of the elastic interaction between the two dislocations has only a small impact on the value of the interaction energy between the O atom and the dislocation.

To obtain the energy associated with the segregation of oxygen on the dislocation, we need to consider that the O atom is coming from a dilute solid solution. The interaction energy (Eq. \ref{eq:Einter}) is therefore calculated, for all O atomic fractions, by considering the same reference for the O solute energy, $E_{\rm O} - E_{\rm bulk}$, obtained with the $5\times5\times4$ hcp supercell where the O atom can be considered isolated (\cf \S \ref{sec:methods}).

As seen in Fig. \ref{fig:Eint_O_py}, the interaction energy is rather constant for cell heights $l_z\geq2b$, \ie for atomic fraction $x_{\rm O}\leq1/2$. This validates our initial choice of a supercell of height $l_z = 2b$. On the other hand, the interaction energy varies rapidly and becomes positive when the supercell height is $l_z = b$. This situation, which corresponds to a full saturation of the dislocation line by O atoms, is therefore not possible thermodynamically and reflects a strong repulsion between O atoms in first neighbor positions. Assuming that the difference in energy between $l_z = b$ and $l_z = 2b$ is due only to the repulsion between O atoms, we obtain a repulsion energy of $98$\,meV, which compares well with the nearest neighbor interaction energy, 120\,meV, computed in a perfect $5\times5\times4$ hcp supercell. This is consistent with the fact that the octahedral insertion sites are similar whether they are located at the dislocation center or in a perfect hcp crystal.

We see from Fig. \ref{fig:Eint_O_py} that the binding energy of the O atom to the screw dislocation does not exceed 63\,meV. 
As a comparison, the activation energy for O diffusion in bulk Zr is 2.08\,eV \cite{Landolt_vol26}. 
The binding energy is therefore too low to lead to any relevant segregation at temperatures where O solutes are mobile.
Considering a simple mean-field thermodynamic model \cite{Ventelon2015} with a constant segregation energy 
$E^{\rm seg}=63$\,meV, the atomic fraction of oxygen segregated in the dislocation core is given by
$x_{\rm O}^{\rm d} = x_{\rm O}^{0} \exp{\left( E^{\rm seg}/kT \right)} / \left[ 1 + x_{\rm O}^0 \exp{\left( E^{\rm seg}/kT \right)} \right]$.
For an oxygen nominal concentration $x_{\rm O}^0=0.01$ typical of zirconium industrial alloys, 
the concentration of O in the dislocation core is only 0.09 at 300\,K and 0.03 at 600\,K.
These values  should be considered as upper approximations as the repulsive interaction energy 
between first nearest neighbour O atoms has been neglected in the segregation energy.


\section{Oxygen hardening}

\begin{figure*}[!bth]
	\begin{center}
	\small
	\begin{tabular}[]{cccccccccc}
		\includegraphics[width=0.085\linewidth]{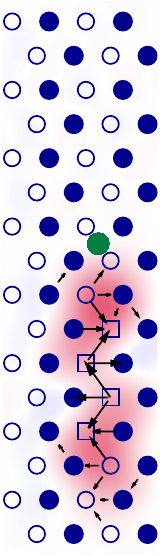}	& 
		\includegraphics[width=0.085\linewidth]{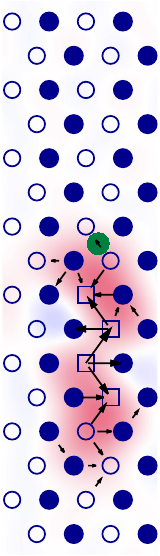} &
		\includegraphics[width=0.085\linewidth]{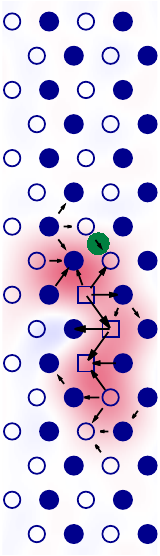}	&
		\includegraphics[width=0.085\linewidth]{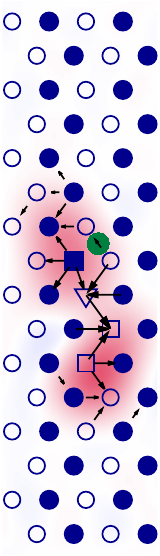}	&
		\includegraphics[width=0.085\linewidth]{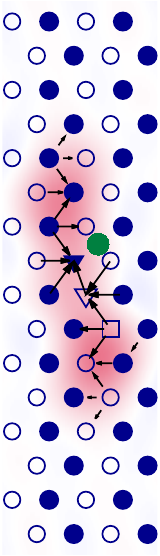} &
		\includegraphics[width=0.085\linewidth]{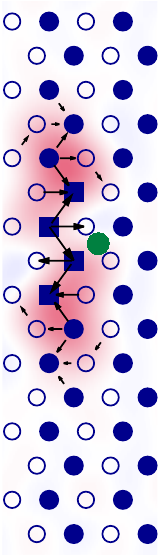}	&
		\includegraphics[width=0.085\linewidth]{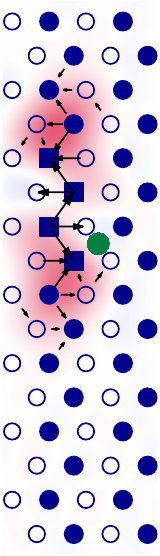} &
		\includegraphics[width=0.085\linewidth]{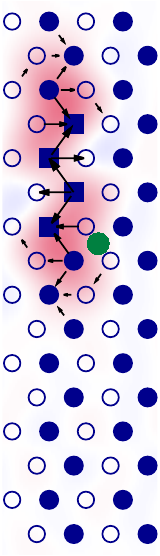}	&
		\includegraphics[width=0.085\linewidth]{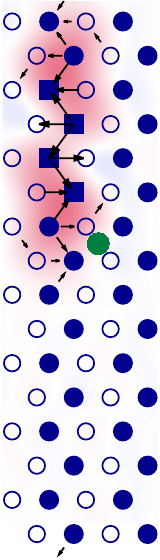}
		\\
		a) $8$\,meV & b) $19$\,meV & c) $54$\,meV & d) $74$\,meV & e) $44$\,meV & f) $57$\,meV & g) $29$\,meV & h) $21$\,meV & i) $16$\,meV 
	\end{tabular}
	\includegraphics[width=0.8\linewidth]{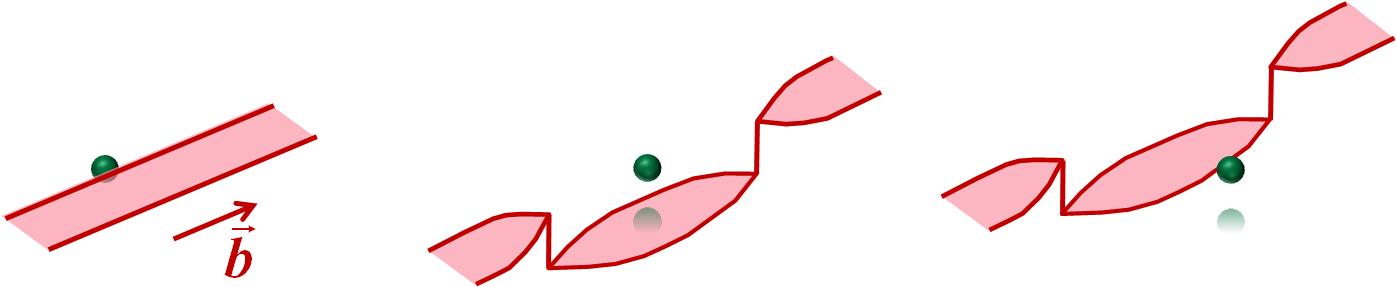}
	\end{center}
	\caption{Dislocation cross-slip induced by an oxygen atom. 
	The oxygen position is indicated by a green circle.
	The O-dislocation interaction energy $E^{\rm int}$, considering the dislocation prismatic configuration as a reference,
	is given below each configuration. For configurations c and d, because of the core reconstruction, 
	part of this interaction energy arises from a variation of the elastic interaction between the two dislocations
	composing the dipole.
	The drawing below illustrates the creation of the two jogs on the screw dislocation induced by the local cross-slip event.}
	\label{fig:cross_slip}
\end{figure*}

We have shown above that the interaction of oxygen atoms with a screw dislocation is mainly repulsive. When the O atom is located out of the dislocation glide plane, the repulsion is small ($E^{\rm int} \lesssim 60$\,meV) and can only lead to a marginal contribution to the hardening observed experimentally. On the other hand, when the O atom belongs to the dislocation glide plane, the repulsion is stronger and may induce a cross-slip of the dislocation, as seen in Fig. \ref{fig:dislo_O_prism}. Using the relaxed configurations obtained for the different positions of the O solute atom, we have reconstructed in Fig. \ref{fig:cross_slip} the path that a screw dislocation would follow upon crossing an O atom. The most striking result is that the dislocation bypasses the solute atom by double cross-slip through the pyramidal plane (Fig. \ref{fig:cross_slip}d-e). The ease of cross-slip arises from the small energy barrier between the dislocation prismatic ground state and its nonplanar metastable state \cite{Chaari2014,Chaari2014a,Clouet2015}. This solute bypass mechanism by dislocation cross-slip is consistent with the experimental observation that oxygen addition favors cross-slip and promotes non-prismatic slip \cite{Baldwin1968}.

Along the bypass mechanism, the interaction energy with the O atom does not exceed 80\,meV, a value just slightly higher than when the O atom is not in the glide plane. However we should keep in mind that the double cross-slip seen here will be localized along the dislocation line and will therefore lead to the creation of a pair of jogs as illustrated in the lower part of Fig. \ref{fig:cross_slip}. DFT calculations are too expensive computationally to include such jogs, but we know from crystallography that they will be of edge character and will therefore move conservatively only along the dislocation line and not in the gliding direction, thus producing a lattice friction. 
Such jogs can also interact with the surrounding solute atoms, thus leading to an elastic interaction between the jogged screw dislocation and the O atoms.
We believe that the resistance due to these jogs on the glide of screw dislocations is at the origin of the extrinsic lattice friction measured experimentally in zirconium alloys containing oxygen. 
This scenario has also been proposed to explain the O hardening in Ti \cite{Yu2015}.
In the case of Zr alloys, it agrees with the analysis of the critical resolved shear stresses and activation volumes
measured for various O contents  \cite{Mills1968,Akhtar1975b,Soo1968,Ardell1966,Moon2006,Morrow2013,Morrow2016}.

Both Mills and Craig \cite{Mills1968} and Akhtar \cite{Akhtar1975b} observed during tensile experiments on Zr single crystals an athermal regime between 600 and 800\,K, followed by a thermal activation of the yield stress with an activation energy corresponding to self diffusion. Mills and Craig \cite{Mills1968} applied the model of Hirsch and Warrington \cite{Hirsch1961} and attributed the athermal regime to the drag of sessile jogs on screw dislocations. 
They assumed that these jogs resulted from the dislocation interaction with O atoms, 
an assumption in agreement with the cross-slip interaction mechanism evidenced in the present \abinitio calculations.
When the temperature becomes high enough to allow for vacancy diffusion, the motion of the jogged screw dislocation is controlled by the vacancy flux at the jogs and the yield stress becomes thermally activated. Soo and Higgings \cite{Soo1968} also conclude from their tensile experiments on single crystals that slip is controlled by the non-conservative motion of jogged screw dislocations above 300\,K, and thus at a lower temperature than in the previous works.
The height of the superjogs estimated from the experimental data is between 1 and 2 atomic distances, consistent with the double cross-slip between adjacent prismatic planes observed here. We note that Ardell \cite{Ardell1966} and Mills \etal \cite{Moon2006,Morrow2013,Morrow2016} also proposed that creep in zirconium alloys is controlled by the motion of jogged screw dislocations through vacancy diffusion \cite{Barrett1965}.

In agreement with these models, Bailey \cite{Bailey1962} observed heavily jogged screw dislocations with post-mortem transmission electron microscopy (TEM) in strained commercial grade zirconium alloys (0.1\,wt.\% O and N) at room temperature. The ``wavyness'' of the screw dislocation, a feature of their jog content, increases with the plastic deformation. The same TEM observations performed in a pure Zr alloy (0.03\,wt.\%O) showed a tangled dislocation microstructure, with no preferential elongation of the dislocations along their screw orientation. Dislocation dipoles resulting from the motion of jogged screw dislocations with a jog height larger than an atomic distance were also observed. Bailey proposed that the jogs created on the screw dislocations were mobile and could therefore condensate to form superjogs on the screw dislocations, but that O atoms decrease their mobility, thus explaining why heavily jogged screw dislocations are observed in the less pure Zr alloy.
Such superjogs were observed by Caillard \etal \cite{Caillard2015b} in  in situ TEM straining experiments. These authors did not support jog drag controlled by vacancy diffusion, but proposed instead an unzipping process corresponding to a conservative glide of the superjogs along the dislocations. 

\begin{figure}[!bh]
	\begin{center}
	\small
	\begin{tabular}[]{cc}
		\hfill
		\includegraphics[width=0.22\linewidth]{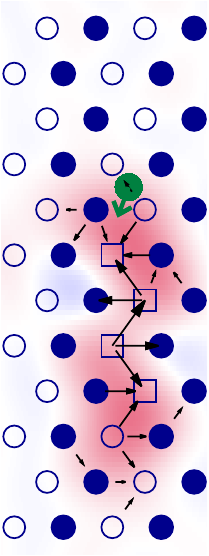} 
		\hfill
		\includegraphics[width=0.22\linewidth]{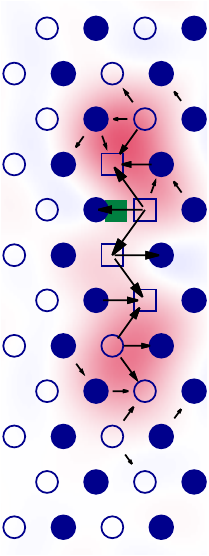} 
		\hfill
		&
		\hfill
		\includegraphics[width=0.22\linewidth]{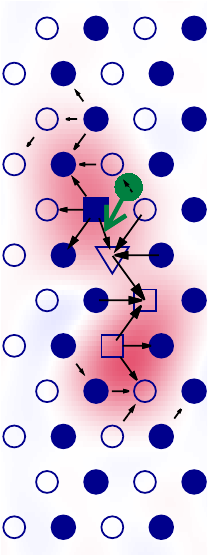}
		\hfill
		\includegraphics[width=0.22\linewidth]{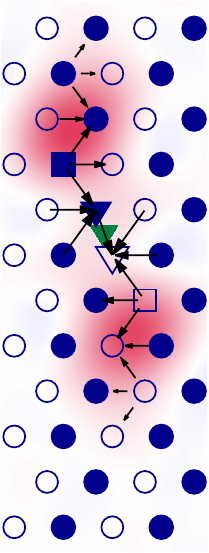}
		\hfill
		\\
		a) 19\,meV \hfill 200\,meV \hfill
		& 
		b) 74\,meV \hfill $-49$\,meV \hfill
	\end{tabular}
	\end{center}
	\caption{Possible shuffling of the O atom when interacting with a screw dislocation. 
	The O atom migrates to an octahedral site created by
	a) the prismatic and b) the pyramidal spreading of the dislocation. }
	\label{fig:shuffle}
\end{figure}

In the above scenario, the interstitial atom remains immobile. However, as illustrated in Fig. \ref{fig:shuffle}a, upon first contact with the screw dislocation, the O atom could possibly jump to one of the new octahedral sites $O_{\rm P}$
created by the prismatic stacking fault. The dislocation would then glide forward to reach the configuration shown in Figs. \ref{fig:dislo_O_prism}g-i and reproduced on the right of Fig. \ref{fig:shuffle}a. This configuration is however of higher energy than any of the cross-slipped configurations and shuffling should therefore operate only if the corresponding excess energy is lower 
than the energy necessary to create the jog pair by double cross-slip. We should note however that bypass by cross-slip is only possible for screw dislocations. Non-screw dislocations, which cannot change their glide plane, will therefore probably interact with the O atoms in their prismatic dissociation plane through such a shuffling mechanism, thus leading to a localized repulsive interaction.
But this shuffling mechanism does not a priori discriminate between screw and non-screw dislocations and can therefore not fully account for the lattice friction induced by O atoms primarily on the screw dislocations.

A second possible shuffling mechanism is available after the first cross-slip event,
when the dislocation has a non-planar core spread in a pyramidal and two adjacent prismatic planes 
(Fig. \ref{fig:cross_slip}d and e).
The O atom can then migrate to the octahedral site $O_{\pi}$ created in the pyramidal stacking fault
(Fig. \ref{fig:shuffle}b).
The final state corresponds to the attractive configuration discussed in the previous section
and has a lower energy than any other configurations. 
However, preliminary NEB calculations (atomic forces converged only to 0.4\,eV\,\AA$^{-1}$)
indicate that the activation energy necessary for the O migration is high ($\sim$1\,eV), thus excluding also this second shuffling mechanism. 

\section{Conclusion}

\Abinitio calculations show that a repulsive interaction exists 
between O atoms and \hkl<a> screw dislocations when the O atoms 
are located in the dislocation habit plane.  
This repulsion results from the destruction 
of the octahedral interstitial sites by the stacking fault associated with the dislocation dissociation.  
This is true both for the dislocation ground state
dissociated in the prismatic plane and for the metastable states
partially or totally dissociated in the pyramidal planes,
since both the prismatic and the pyramidal stacking faults destroy 
the O insertion sites. 
As a consequence of this repulsive interaction, the screw dislocation 
cross-slips to restore the octahedral insertion site. 
This cross-slip event induced by the oxygen atom 
will create two jogs on the screw dislocation, which probably explains 
the lattice friction acting against screw dislocation glide in zirconium alloys 
containing oxygen.  Such a scenario is in agreement with the hardening and creep
models proposed to rationalize tensile tests and creep experiments in Zr alloys, 
although the dynamics of the jogs along the dislocation line and whether they will annihilate or coalesce after unpinning from O atoms remains to be explored. To this end, the bypass mechanism must be simulated in three dimensions with the dislocation in presence of a representative distribution of O atoms. DFT calculations are obviously too expensive to perform such calculations, but the present work can serve in the future as input to develop approximate but reliable models of atomic interactions.

\vspace{0.5cm}
\linespread{1}
\small

\textbf{Acknowledgements} -
This work was performed using HPC resources from GENCI-CINES and -TGCC (Grants 2015-096847).
The author also acknowledge PRACE for access to the Curie resources based in France at TGCC 
(project PlasTitZir).
DR acknowledges support from LABEX iMUST (ANR-10-LABX-0064) of Université de Lyon 
(program ``Investissements d'Avenir'', ANR-11-IDEX-0007)

\section*{References}
\bibliographystyle{ActaMatnew-2.bst}
\bibliography{chaari2017}

\end{document}